\documentstyle[multicol,aps,prb,epsf]{revtex}

\newcommand{\Vec}[1]{\mbox{\boldmath$#1$}}
\newcommand{\GVec}[1]{\mbox{\boldmath$#1$}}

\begin{document}

\draft

\title{Integer quantum Hall effect and Hofstadter's butterfly spectra\\
in three-dimensional metals in external periodic modulations}
\author{Mikito Koshino\cite{addr}, Hideo Aoki}
\address{Department of Physics, University of Tokyo, Hongo, Tokyo
113-0033, Japan}
\date{\today}

\maketitle

\begin{abstract}
We propose that Hofstadter's butterfly accompanied by quantum Hall effect 
that is similar to those predicted to occur in 3D tight-binding systems 
by Koshino {\it et al.} [Phys. Rev. Lett. {\bf 86}, 1062 (2001)] 
can be realized in an entirely different system --- 3D metals applied with
weak external periodic modulations (e.g., acoustic waves).  
Namely, an effect of two periodic potentials 
interferes with Landau's quantization due to an 
applied magnetic field $\Vec{B}$, resulting generally 
in fractal energy gaps as a function of the tilting angle of $\Vec{B}$, 
for which the accompanying quantized Hall tensors 
are computed.  The phenomenon arises from the fact that, 
while the present system has a different physical origin for the butterfly 
from the 3D tight-binding systems, the mathematical forms are 
remarkably equivalent. 
\end{abstract}

%\newpage

\begin{multicols}{2}
\narrowtext

{\it Introduction}
In two-dimensional(2D) periodic systems in magnetic fields, 
it is well-known that the interplay of Bragg's
reflection and Landau's quantization gives rise to a fractal series of
energy gaps, which is called Hofstadter's butterfly\cite{Hofs}.
Theoretically, the self-similar spectrum emerges both in 
the strong potential limit\cite{Hofs} (tight-binding system) and
in the weak potential limit\cite{Rauh} (Landau-quantized system),
where two cases relate to each other in their mathematical expressions.
When the Fermi energy is within a gap in the butterfly
the system exhibits the quantum Hall effect, where the Hall conductivity
is expressed as a topological quantum number for each energy
gap\cite{TKNN}. In recent experiments, a sigunature 
of Hofstadter's butterfly has been reported
in the Hall conductivity measurement for 
semiconductor superlattices\cite{Albr}.

While the butterfly spectra had been known to be peculiar to 2D,
we have previously found that {\it three-dimensional(3D)
tight-binding lattice} can have a fractal energy spectra 
like Hofstadter's when the system is anisotropic (quasi-1D),
where the 3D butterfly is shown to be a genuinly 3D effect 
rather than a remnant of a 2D butterfly\cite{Kosh}.  
This leads to an interesting transport phenomenon --- 
quantum Hall effect (QHE) in three dimensions.
According to the general argument\cite{Halp,Mont,Kohm}, 
QHE occurs even in 3D {\it provided} that there is an energy gap 
and that the Fermi energy lies in the gap, where the three components of the
Hall conductivity tensor should be individually quantized.

Then an intriguing question arises: can we have 
a counterpart in the opposite limit, {\it 3D weakly modulated systems}?
Here we investigate the energy spectra in 3D continuous systems with
weak periodic potentials, and show that
a modulation composed of two plane waves (such as two acoustic waves) 
gives rise to Hofstadter's butterfly, 
although the mechanism is quite distinct from one
in the tight-binding system. 
We also find a clear mathematical relationship
connecting two limiting cases similarly to 2D,
but we will see that three-dimensionality is essential 
throughout.

The problem has another interesting aspect, 
i.e., the QHE topological numbers.  
Here we have calculated the Hall conductivity in 3D weak
potential systems, and find that the expression is
remarkably equivalent to that for the corresponding energy gap 
in the tight-binding system\cite{Kosh}, 
which is unlike the relationship in 2D 
between the limits of weak $\leftrightarrow$ strong potentials.

The present experimental setup (just an 
ordinary 3D metal with two acoustic waves applied) 
is so simple that this can be 
a strong candidate for experimentally detecting the 3D QHE 
and Hofstadter's butterfly.   We shall estimate the 
required magnetic field and acoustic wave  length, etc, 
for a possible realization in a semimetal 
subjected.

{\it Formulation: One modulation} 
We first look at how a single modulation interferes with 
Landau's quantization due to a magnetic field $\Vec{B}$
in a 3D uniform electron gas.  The Hamiltonian is simply 
${\mathcal H} = (1/2m)(\Vec{p}+e\Vec{A})^2 + U(\Vec{r})$, 
where $U(\Vec{r})$ is a perturbative periodic potential,
and the vector potential is taken as $\Vec{A} = (0,Bx,0)$
for $\Vec{B} = (0,0,B).$
The eigenstates in the absence of $U(\Vec{r})$ are, as usual,
\begin{eqnarray}
|n,k_y,k_z\rangle &=& N_n \exp (ik_y y + ik_z z) 
\Psi_n(x/l + k_y l), \nonumber\\
\Psi_n(z) &=& e^{-z^2/2}H_n(z),
\end{eqnarray}
where $H_n$ is the Hermite polynomial with the Landau index $n$,
$N_n$ a normalization factor, and $l = \sqrt{\hbar/(eB)}$
is the magnetic length ($\sim 80$ \AA for $B=10$ T). 
The eigenenergy is
$E_{n,k_z} = \left(n+1/2\right)\hbar\omega_c + \hbar^2k_z^2/(2m)$
with the electron mass $m$ and the cyclotron frequency $\omega_c = eB/m$, 
which suggests that a 3D uniform electron gas becomes 1D in that 
the motion along the magnetic field remains free.  
So, if we have an external modulation on top,
we immediately expect energy gaps to emerge in the 1D band. 

\begin{figure}
%h=here, t=top, b=bottom, p=separate figure page
\begin{center}\leavevmode
\leavevmode\epsfxsize=85mm \epsfbox{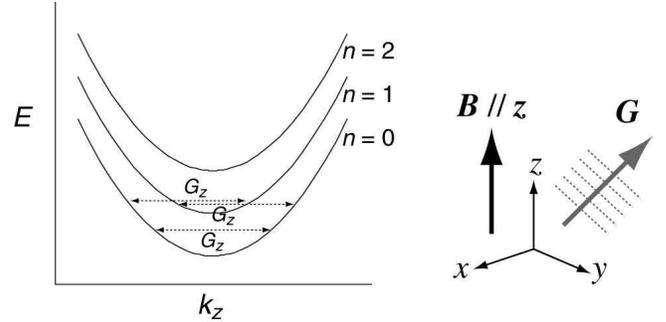}
\caption{The energy dispersion plotted against the wavenumber
along $z$ ($\parallel \Vec{B}$) in a continuous 3D metal 
in a magnetic field $\Vec{B}$ applied in a general direction (inset).  
Dashed arrows indicate where the modulation having 
$G_z$ (the $z$-component) makes the dispersion gapped.
} 
\label{weak_pot}
\end{center}
\end{figure}

When we apply a single periodic potential 
$U(\Vec{r}) = U_0 \cos (\Vec{G}\cdot\Vec{r})$ in 3D, 
a mixing between the states $|n,k_y,k_z\rangle$ and
$|n',k_y \pm M G_y,k_z \pm M G_z\rangle$
gives rise to energy gaps in the $M$-th order perturbation.
The energy gaps due to the mixing between higher Landau bands
will generally tend to be hidden by other bands, so we focus on the
lowest two levels. Namely, energy gaps should open 
within the lowest Landau level ($n=n'=0$) 
and between the lowest and second levels ($n=0, n'=1$) 
(see, Fig. \ref{weak_pot}).
The corresponding matrix elements in the first order 
perturbation are
\begin{eqnarray}
\langle 0,&k_y& \pm G_y,k_z \pm G_z | U(\Vec{r})| 0,k_y,k_z\rangle
\nonumber \\
&=& U_0 e^{-G_{\perp}^2l^2/4}
e^{\mp i\left( k_y \pm \frac{G_y}{2}\right)G_{x} l^2}, \label{Vkk}\\
\langle 1,&k_y& \pm G_y,k_z \pm G_z | U(\Vec{r})| 0,k_y,k_z\rangle
\nonumber \\
&=& U_0 e^{-G_{\perp}^2l^2/4}
e^{\mp i\left( k_y \pm \frac{G_y}{2}\right)G_{x} l^2}
\times i(G_x-iG_y)l/\sqrt{2}
\end{eqnarray}
where $G_{\perp} \equiv \sqrt{G_x^2+G_y^2}$.
Since the matrix elements scale as 
$\exp(-G_{\perp}^2l^2/4)$,
the magnitude of the matrix elements, hence 
the size of the energy gap, becomes significant 
when $G l \lesssim 1$ 
(i.e., $\exp(-G_{\perp}^2l^2/4)$ vanishes 
like $\exp(-{\rm const.}/B)$ for $B \rightarrow 0$).
The condition $G l \lesssim 1$ also suggests that
the main energy gaps open 
for  $E \lesssim \hbar\omega_c$, i.e., around the lowest or 
the second lowest Landau levels.

In 3D systems, the Hall conductivity has three components, 
$\GVec{\sigma} \equiv (\sigma_{yz},\sigma_{zx},\sigma_{xy})$,
where the Hall current in an electric field $\Vec{E}$ is given by 
$\Vec{j} = \GVec{\sigma} \times \Vec{E}$.
We can immediately calculate $\Vec{\sigma}$ for the present system
when the Fermi energy $E_F$ is in each of the energy gaps.
If we consider generally the $M$-th order gap ($M=1,2,...$) 
within the lowest Landau level ($n=n'=0$),
the number of states below $E_F$ is
\begin{equation}
N_F = \frac{M G_z}{2\pi/L_z} \times \frac{L_xL_y}{2\pi l^2}
= \frac{eV}{2\pi h}M\Vec{G}\cdot\Vec{B},
\end{equation}
where $V$ is the system volume.
From Widom-St\v{r}eda's formula\cite{Wido,Stre},
$\GVec{\sigma} = -(e/V)(\partial N_F / \partial \Vec{B})$,
we readily obtain
\begin{equation}
\GVec{\sigma} = -\frac{e^2}{2\pi h}M\Vec{G} \,\,\,\,\,\,\,
(n = n'= 0).\label{sigma_LLL}
\end{equation}
Similarly, we have for the inter-Landau gap 
\begin{equation}
\GVec{\sigma} = -\frac{e^2}{2\pi h}2M\Vec{G} \,\,\,\,\,\,\,
(n = 0, n'= 1).\label{sigma_2LL}
\end{equation}

A note is due here.  3D continuous systems 
having a one-dimensional modulation in magnetic fields 
have been investigated for a long time 
in terms of density-wave (DW) instabilities.  
Specifically, Halperin\cite{Halp} has calculated 
the Hall conductivity when the system is in a DW state,
which corresponds to our calculation
for the energy gap within the lowest Landau level.
The {\it inter-Landau level} gap formation is 
new to the best of our knowledge.

{\it Two modulations} 
Now, if we superpose two modulations 
$U_1(\Vec{r}) = U_1\cos ({\Vec{G}_1\cdot\Vec{r}})$ and 
$U_2(\Vec{r}) = U_2\cos ({\Vec{G}_2\cdot\Vec{r}})$ 
having different wavevectors,
the discussion above can be extended, where 
we can substitute $M \Vec{G}$ with $M \Vec{G}_1 + N \Vec{G}_2$
for the energy gap corresponding to
$M$th and $N$th-order perturbations in $U_1$ and $U_2$, respectively.
The Hall conductivity for the gap with the order $M,N$
in the lowest Landau band thus becomes
\begin{equation}
\GVec{\sigma} = -\frac{e^2}{2\pi h}(M\Vec{G}_1 + N\Vec{G}_2)
 \,\,\,\,\,\,\,(n = 0, n'= 0),
\label{sigma_weak}
\end{equation}
where $(M,N)$ becomes $(2M,2N)$ for the $(M,N)$-th
inter-Landau level gap.

We can immediately realize that the spectrum should be 
sensitively affected by the commensurability of 
the ratio between the $z$-components, $G_{1z}$ 
and $G_{2z}$, since, for instance if the ratio is irrational 
$M G_{1z} + N G_{2z}$ can take continuous values and 
the spectrum should have gaps everywhere in the energy axis.  
If the magnetic field is {\it rotated} in the plane
spanned by $\Vec{G}_1$ and $\Vec{G}_2$, 
the energy spectrum is then expected to have a fractal structure
since the commensurability between $G_{1z}$ and $G_{2z}$ varies in a 
complicated manner.  

\begin{figure}
\begin{center}\leavevmode
\leavevmode\epsfxsize=85mm \epsfbox{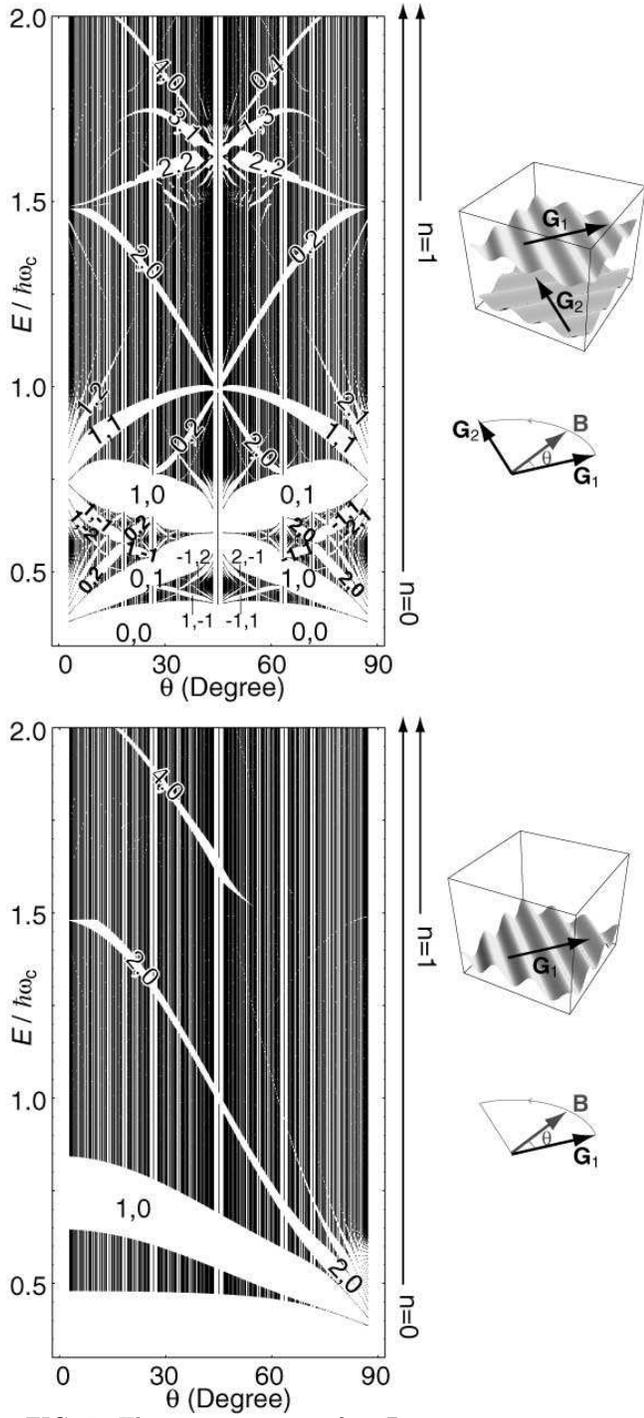}
\caption{
The energy spectra 
for 3D continuous systems in two periodic modulations 
having wavevectors $\Vec{G}_1,\Vec{G}_2$ (top), 
along with the plot for one modulation $\Vec{G}_1$ (bottom),
for $|G_i|=\sqrt{2}/l$, $\Vec{G}_1 \perp \Vec{G}_2$, 
and the amplitudes of the periodic potentials of $0.2\hbar\omega_c$.
The horizontal axis is
the angle of $\Vec{B}$ rotated in the plane
spanned by $\Vec{G}_1$ and $\Vec{G}_2$.
A pair of integers $(M,N)$ attached to each energy gap
represents the Hall integers, eq.(\ref{sigma_weak}).
The spectrum has been plotted for
rational angles, $\tan \theta = p/q$ ($p,q$: mutually prime integers), 
with an energy cut-off at $4.5\hbar\omega_c$.
The arrows on the right indicate the energy regions for the 
Landau bands in the un-modulated case.  
} 
\label{G1v02}
\end{center}
\end{figure}

The spectrum plotted in Fig.\ref{G1v02} against 
the tilting angle of $\Vec{B}$ obtained numerically, has indeed 
butterfly-like structure in the lowest Landau band
as in Hofstadter's spectum.
The Hall integers $M,N$ have been
calculated rigorously for each gap, by tracing the number of states
below the gaps for the tilted field.
In higher energy region where two Landau levels 
overlap ($E > (3/2)\hbar\omega_c$) 
we see the gap structure due to the inter-Landau level mixing
(the gap labeled with $M,N = 4,0 (0,4)$ and 2,2 in the figure), 
as well as the intra-level gaps which are visible
only when gaps within $n=0$ and $n=1$ happen to overlap.  
In the figure, the gap 3,1 (1,3) is a composite of
2,1 (1,2) in the lowest level and 1,0 (0,1) in the second.

{\it Correspondence with the strong potential case} 
As mentioned, we have previously found a butterfly-like spectra
for the first time in periodic 3D systems
which is modelled 
by the anisotropic (quasi-1D) {\it tight-binding} lattice.
There we have shown that the period along the most 
conductive direction ($x$) is {\it not} responsible for
the emergence of the 3D butterfly, 
while other two periods (along $y,z$) are relevant.
Thus the two (present and the previous) cases 
are both {\it 3D system with two periods}
and the only difference is the strength of the 
periodic potential.
One might then be tempted to think that the two butterfly spectra
cross over to each other
when the amplitude of the periodic modulation is increased or decreased.  
However, as shown in the following,
they have distinct physical origins, residing in different limits, 
and the resemblance in the spectra comes from a
mathematical relationship.

To show this, let us concentrate on weak modulations with 
$G_1, G_2 \lesssim 1/l$  for the lowest Landau level 
(i.e., low-energy region of the spectrum).  
To clarify the relationship, we concentrate on a case
where $\Vec{B}$ lies on the plane spanned by $\Vec{G}_1,\Vec{G}_2$.  
If we take $\Vec{G}_1$ and $\Vec{G}_2$
on $zx$-plane (with $G_{1y}=G_{2y}=0$) so that $k_y$ is conserved, 
Schr\"{o}dinger's equation becomes 
\begin{equation}
 \frac{\hbar^2k_z^2}{2m}  \psi(k_z) 
+ \sum_{k'_z}  U_{k_z,k'_z} \psi(k'_z) = E \psi(k_z), 
\end{equation}
where $\psi(k_z)$ is the amplitude of 
$|0, k_y, k_z\rangle$ with $k_y$ being constant, 
$U_{k_z,k'_z}$ is eq.(\ref{Vkk}) summed over 
$U_1$ and $U_2$ with $G_y = 0$, and is a function of $k_y$. 
By Fourier-transforming with respect to $z$ we obtain
\begin{eqnarray}
 -\frac{\hbar^2}{2m} \frac{\partial^2}{\partial z^2}  \psi(z) 
&& + \tilde{U_1} \cos (G_{1z} z - k_yG_{1x}l^2) \psi(z) \nonumber\\
&& + \tilde{U_2} \cos (G_{2z} z - k_yG_{2x}l^2) \psi(z)
= E \psi(z),\label{eq_weak} 
\end{eqnarray}
where $\tilde{U_j} =  U_j e^{-G_{j\perp}^2l^2/4}$
with $G_{j\perp} \equiv \sqrt{G_{jx}^2+G_{jy}^2}$ ($j = 1,2$).

Now let us recapitulate how a butterfly spectrum arose in our original 
model on a 3D tight-binding lattice\cite{Kosh}.
We consider the 3D orthorhombic lattice
with the nearest-neighbor transfers $t_x,t_y,t_z$ along
$x,y,z$, respectively.
We assume here that the magnetic field $\Vec{B}$ is applied
parallel to $yz$-plane ($B_x = 0$), 
which corresponds to the assumption made in the weak potential case
that $\Vec{G}_1, \Vec{G}_2, \Vec{B}$ are co-planar.
We also assume that the system is quasi-1D ($t_x \gg t_y,t_z$),
and apply the effective mass approximation to the 
motion along the conductive direction $x$,
to obtain the low-energy spectra ($\lesssim t_y$ or $t_z$ from the bottom).
In a gauge 
$\Vec{A} = (0, B_z x, - B_y x)$ we can take 
the basis $\Psi(\Vec{r}) = e^{ik_y y + i k_z z}\psi(x)$, 
and Schr\"{o}dinger's equation becomes 
one-dimensional\cite{Mont,Kosh}:
\begin{eqnarray}
 -\frac{\hbar^2}{2m} \frac{\partial^2}{\partial x^2}  \psi(x) 
&& - 2t_y \cos \left(\frac{eB_z b}{\hbar} x + k_y b \right) \psi(x) \nonumber\\
&& \hspace{-20mm}
- 2t_z \cos \left(-\frac{eB_y c}{\hbar} x + k_z c \right) \psi(x) 
= E \psi(x)\label{eq_strong}, 
\end{eqnarray}
where $b,c$ are the lattice constants along $y,z$, respectively.
Now we can see that eq.(\ref{eq_weak}) and (\ref{eq_strong})
have the identifal form: 1D equation with a double period.
We should recall that Hofstadter's butterfly emerges generally 
in the spectrum of a doubly periodic 1D system when the 
spectrum is plotted against the ratio of the two periods\cite{Kosh}.
In each of two equations, the ratio changes continuously
when the magnetic field is rotated relative to the periodic potential,
which is why Hofstadter's butterfly arises 
against the tilting angle in both cases.

From the correspondence we can also note 
that $\tilde{U_1}, \tilde{U_2}$ in eq.(\ref{eq_weak}) 
correspond to $t_y, t_z$ in (\ref{eq_strong}), respectively, 
but their physical meaning is opposite in the following sense.  
Namely, 
$\tilde{U_1}, \tilde{U_2} \rightarrow 0$ describes the limit 
of weak periodic potential, 
while $t_y, t_z \rightarrow 0$ the strong limit.
The equations become purely 1D in both limits,
where the system reduces to `Landau tubes' 
(1D motion along $\Vec{B}$ and the cyclotron motion on the 
plane $\perp \Vec{B}$) 
in the former, while in the latter wires along $x$ 
confined by the strong potential on $yz$ plane.
On the other hand, 
if the perturbations ($\tilde{U_1}, \tilde{U_2}$  or $t_y, t_z$)
are too large, the mixing between different 1D channels
(i.e., different Landau levels in the former,
different bound modes on $yz$ plane in the latter) becomes so strong 
that the one-band approximation breaks down.
We thus conclude that the two butterfly spectra 
in fact correspond to opposite limits (weak/strong potential).

We can also establish a relationship 
between the Hall conductivities between the two cases.
In the strong potential the Hall conductivity for the gap with 
$M$-th in $t_y$ and the $N$-th in $t_z$ in the tight-binding band 
can be obtained similarly as\cite{Kosh}
\begin{equation}
\GVec{\sigma} = 
-\frac{e^2}{h}\left(0,\frac{N}{b},\frac{M}{c}\right).
\end{equation}
Since the primitive reciprocal lattice vectors are
$\Vec{G}_1 = (2\pi/c) \hat{\Vec{e}}_z$ and
$\Vec{G}_2 = (2\pi/b) \hat{\Vec{e}}_y$, 
this coincides with
the corresponding expression in the weak potential, 
eq.(\ref{sigma_weak}).  
This is rather remarkable, since 
the corresponding wave functions in the two cases 
have totally different spatial behaviors, while 
they carry the identical Hall current.  
This contrasts with the 2D case, where
the Hall integers are different between 
the corresponding gaps in the weak- and strong-potential limits.

% localization

Some notes on the quantum Hall effect are due here.   
First, the non-monotonic behavior in 
the Hall conductivity as $E_F$ is increased 
(see, Fig. \ref{G1v02}) 
should be a hallmark of the 3D butterfly spectrum 
as in the 2D butterfly.  
As for the quantum Hall plateaus, 
we have to consider the localization effect 
in the presence of disorder.  
While this is an interesting future problem, 
we speculate that each subband would evolve into
the localized and the extended states with mobility edges 
since we have a 3D system here, 
and that the Hall conductivity would be constant as long as $E_F$
stays in the localized region. 
For the usual 2D butterfly, a numerical study for a dirty, 
finite system\cite{2Dloc} 
shows that we still have a nonmonotonic behavior as
a sign for the butterfly when the disorder is not too strong,
so we expect a similar behavior in the disordered 3D butterfly as well.

Second, while our calculation of the Hall conductivity
is based on the bulk description,
a finite system has edge states in the bulk gaps,
which also contribute to the Hall current.  
We have previously shown\cite{KHA} that 
the Hall conductance in a finite 3D system is still quantized, 
in the 3D QHE condition, when we take into account both of the contributions 
from the surface current (which we have called the wrapping 
current) and from the bulk current, where the quantized values 
exactly coincides with those for the infinite system. 
The argument is quite general, and applies to the present problem as well.

{\it Experimental feasibility} 
Finally let us comment on the experimental feasibility
for the weakly modulated QHE system proposed here.
To observe the QHE, $E_F$ have to reside in an energy gap,
which appears only in the lowest Landau levels.
So we require the situation where only a few
levels are occupied in 3D (the quantum limit),
for which we need a large magnetic field or a small electron 
concentration.  
We expect that semimetals should be suitable for the latter condition.  
We can estimate the required magnetic field for the quantum limit in bismuth, 
for example, to be $B\gtrsim 10$T, which is quite modest.  
More stringent is the condition for the modulation, since 
the wave length should be such that the Fermi level is in the 
gap created by the modulation.
If we take the acoustic wave for the periodic potential
and plug in the velocity of sound for Bi, 
we have $f \sim 100$GHz for $B = 10$T, 
which is rather high,
although the required frequency
($\propto$ Fermi wavenumber)
decreases with $B$ like $f \propto 1/B$.
The coupling of the electrons with the acoustic 
wave, on the other hand, 
is dominated by the piezoelectric coefficient of the material, 
so compounds may be advantageous in this respect.  
An obvious advantage of the externally modulated system 
proposed here over the tight-binding system is
that we can change the wavelength of the external modulation 
at our disposal.  So we expect that we can find a wider possibility for
observing Hofstadter's butterfly with the 3D QHE.

% conclusion
%In conclusion, we investigated the energy spectra
%and the quantum Hall effect in a continuous 3D system
%with several modulations.
%There we have shown that 
%(i) a one-dimensional modulation gives rise to a series of
%energy gaps characterized with 
%one component of the quantized Hall tensor,
%and (ii) a two-dimensional modulation causes {\it Hofstadter's butterfly}
%with two components of $\sigma_{ij}$ being quantized independently.

\end{multicols}
\end{document}